\documentclass[fleqn,usenatbib]{mnras}
\usepackage[T1]{fontenc}
\usepackage[T1]{fontenc}
\usepackage[latin9]{inputenc}
\usepackage{graphicx}	
\usepackage{amsmath}	
\usepackage{amssymb}	
\usepackage[dvipsnames, table]{xcolor}
\usepackage{multirow}
\usepackage[section]{placeins}
\usepackage{multicol}
\usepackage{upgreek}    
\usepackage{enumitem}   
\setlist[enumerate]{leftmargin=*}   
\setlist[itemize]{leftmargin=*}   

\usepackage{txfonts}

\usepackage{soul}

\usepackage{orcidlink}

\usepackage{xifthen}

\newcommand{\msol}{\,\tr{M}_{\odot}}

\newcommand{\pc}{\,\mathrm{pc}}
\newcommand{\yr}{\,\mathrm{yr}}

\newcommand{\tr}[1]{\textrm{#1}}
\newcommand{\trt}[1]{\textrm{\tiny{#1}}}

\newcommand{\E}[1]{\times\nobreak10^{#1}}

\newcommand{\logten}[1]{\log_{10}\!\lr{#1}}

\newcommand{\lr}[2][]{
    \ifthenelse{\equal{#1}{}}{
        {\left(#2\right)}
    }{
        {\left(#2\right)}^{#1}
    }
}

\newcommand{\lrs}[2][]{
    \ifthenelse{\equal{#1}{}}{
        {\left[#2\right]}
    }{
        {\left[#2\right]}^{#1}
    }
}

\newcommand{\scale}[3][]{
    \ifthenelse{\equal{#1}{}}{
        \lr{ \frac{#2}{#3} }
    }{
        {\lr[#1]{ \frac{#2}{#3} }}
    }
}

\newcommand{\sinpar}[2][]{\sin^{#1}\!\lr{#2}}

\newcommand{\figref}[1]{Fig.~\ref{#1}}

\newcommand{\astropy}{\texttt{Astropy}}
\newcommand{\matplotlib}{\texttt{matplotlib}}
\newcommand{\numpy}{\texttt{NumPy}}
\newcommand{\scipy}{\texttt{SciPy}}
\newcommand{\ipython}{\texttt{ipython}}
\newcommand{\jupyter}{\texttt{jupyter}}

\DeclareRobustCommand\kalepy{\texttt{
    k\kern+0.05em%
    \raisebox{0ex}{$a$}\kern-.10em%
    \textcolor{black}{$l$}\kern-0.0em%
    e\kern+0.04em%
    p\kern-.05em%
    y%
}}

\def\apj{Astrophys. J.}       		        	 	
\def\apjl{Astrophys. J. Lett.}             		
\def\pasj{PASJ}

\def\pasp{PASP}

\def\apjs{Astrophys. J., Suppl. Ser.}            
\def\mnras{Mon. Not. R. Astron. Soc.}        
\def\araa{Annu. Rev. Astron. Astrophys.}  
\def\nat{Nature}              				
\def\aap{Astron. Astrophys.}               		

\newcommand{\fluxunits}{\mathrm{erg/s/cm}^2\mathrm{/Hz}}
\newcommand{\tdrw}{\tau_\trt{DRW}}       
\newcommand{\sfinf}{\mathrm{SF}_\infty}  
\newcommand{\sigmadrw}{\sigma_\trt{DRW}} 
\newcommand{\mean}[1]{\langle{#1}\rangle}
\newcommand{\fedd}{f_\textrm{Edd}}          
\newcommand{\feddsec}{f_{\textrm{Edd},2}}   
\newcommand{\feddi}{f_{\textrm{Edd},i}}   
\newcommand{\pobs}{p}          
\newcommand{\pobsmax}{p_\textrm{max}}   
\newcommand{\fsens}{F_{\nu,\textrm{sens}}}  
\newcommand{\tlens}{\tau_\textrm{lens}}     
\newcommand{\tlensi}{\tau_{\textrm{lens},i}}     
\newcommand{\snr}{\mathrm{SNR}}     
\newcommand{\fnoise}{F_\textrm{noise}}
\newcommand{\incorb}{I_\textrm{orb}}
\newcommand{\incdisk}{I_\textrm{disk}}
\newcommand{\rdust}{R_\textrm{dust}}
\newcommand{\sepa}{a_\textrm{sep}}
\newcommand{\vpm}[3]{$#1^{#2}_{#3}$}
\newcommand{\mvpm}[3]{#1^{#2}_{#3}}

\let\textacute\'
\let\textgrave\`
\newcommand{\cafg}{Claude-Andr\textacute{e} Faucher-Gigu\textgrave{e}re}

\newcommand{\eg}{\textit{e.g.}} 

\sodef\tight{}{-0.025em}{0.2em plus 0.2em}{0.5em plus 0.1em minus 0.1em}

\definecolor{mplblue}{HTML}{1f77b4}
\definecolor{mplorange}{HTML}{ff7f0e}
\definecolor{mplgreen}{HTML}{2ca02c}
\definecolor{mplred}{HTML}{d62728}
\definecolor{mplpurple}{HTML}{9467bd}
\definecolor{mplbrown}{HTML}{8c564b}
\definecolor{mplpink}{HTML}{e377c2}

\definecolor{uband}{HTML}{ee82ee}
\definecolor{bband}{HTML}{0000ff}
\definecolor{vband}{HTML}{008000}
\definecolor{rband}{HTML}{ff0000}
\definecolor{iband}{HTML}{8b0000}

\title[MBH Binary Self-Lensing]{Gravitational Self-Lensing in Populations of Massive Black Hole Binaries}

\author[L.~Z.~Kelley et al.]{
    Luke Zoltan Kelley\,\orcidlink{0000-0002-6625-6450},$^{1,2}$\thanks{lzkelley@northwestern.edu}
    Daniel J. D'Orazio\,\orcidlink{0000-0002-1271-6247},$^{3}$
    Rosanne Di Stefano\,\orcidlink{0000-0003-0972-1376}$^{4,5}$
    \\
    $^{1}$CIERA (\tight{Center For Interdisciplinary Exploration And Research In Astrophysics}), 1800 Sherman Ave, Evanston, IL 60201, USA \\
    $^{2}$Dept.~of Physics \& Astronomy, Northwestern University, 2145 Sheridan Road, Evanston, IL 60208, USA \\
    $^{3}$Niels Bohr International Academy, Niels Bohr Institute, Blegdamsvej 17, 2100 Copenhagen, Denmark \\
    $^{4}$Smithsonian Astrophysical Observatory, 60 Garden Street, Cambridge, MA 02138, USA \\
    $^{5}$Astronomy Department, Harvard University, 60 Garden Street, Cambridge, MA 02138, USA
}

\date{Accepted XXX. Received YYY; in original form ZZZ}
\pubyear{2021}

\begin{document}
\label{firstpage}
\pagerange{\pageref{firstpage}--\pageref{lastpage}}
\maketitle


\begin{abstract}
    The community may be on the verge of detecting low-frequency gravitational waves from massive black hole binaries (MBHBs), but no examples of binary active galactic nuclei (AGN) have been confirmed.  Because MBHBs are intrinsically rare, the most promising detection methods utilize photometric data from all-sky surveys.  Recently \citet{dorazio+distefano-2018} suggested gravitational self-lensing as a method of detecting AGN in close separation binaries.  In this study we calculate the detectability of lensing signatures in realistic populations of simulated MBHBs.  Within our model assumptions, we find that VRO's LSST should be able to detect 10s to 100s of self-lensing binaries, with the rate uncertainty depending primarily on the orientation of AGN disks relative to their binary orbits.  Roughly a quarter of lensing detectable systems should also show detectable Doppler boosting signatures.  If AGN disks tend to be aligned with the orbit, lensing signatures are very nearly achromatic, while in misaligned configurations the bluer optical bands are lensed more than redder ones.   Whether substantial obscuring material (e.g.~a dusty torus) will be present in close binaries remains uncertain, but our estimates suggest that a substantial fraction of systems would still be observable in this case.
\end{abstract}



\section{Introduction}
    
    Binaries of massive black holes (MBHs) are expected to be formed following the merger of their host galaxies \citep{Begelman+1980}.  These binaries are promising sources of low-frequency gravitational waves which will be detectable by pulsar timing arrays \citep{Detweiler-1979, Hellings+Downs-1983, Foster+Backer-1990}.  In particular, the gravitational wave signal from the ensemble of MBH binaries (MBHBs) across the univerise---the stochastic gravitational wave background \citep{Rajagopal+Romani-1995, Wyithe+Loeb-2003, Jaffe+Backer-2003, Sesana+2004}---may be starting to emerge in recent pulsar timing array data \citep{NANOGrav+2020_12p5gwb}.
    
    A wide array of possible electromagnetic signatures of accreting, sub-parsec MBHBs have been proposed \citep[see:][for reviews]{Burke-Spolaor-2013, Bogdanovic-2015, DeRosa+2019}, but no observational examples have been confirmed.  Because MBHBs are intrinsically rare \citep[$\sim 10^{-3} \, \textrm{AGN}^{-1}$; ][]{Kelley+2019}, signatures which can be detected in large surveys, particularly with photometric data alone, are especially promising.  For example, a growing number of candidate MBHBs have been identified based on apparent periodicity in AGN lightcurves \citep{Graham+2015, Charisi+2016, LiuSBHB+2019,ChenDES_SBHBs+2020}    \footnote{c.f.~\citet{Sesana+2018,Kelley+2019}.}.  The Vera Rubin Observatory (VRO) Legacy Survey of Space and Time \citep[LSST;][]{Ivezic+2019_lsst}, which is expected to begin in roughly $2024$, will vastly increase the sample size of time-domain AGN observations and revolutionize both AGN and binary-AGN astrophysics. 
    
    Recently, \citet{dorazio+distefano-2018} considered the photometric signatures of one MBH in a binary gravitationally \mbox{(micro-)lensing} a companion\footnote{an approach also suggested for identifying compact remnants and planets in stellar systems \citep[e.g.][]{Maeder-1973, Mao+Paczynski-1991, Gould+Loeb-1992, Beskin+Tuntsov-2002, Rahvar+2011}.} active galactic nucleus (AGN).  In addition to the rarity of binary AGN, detectable `self-lensing' systems are made even more infrequent by necessitating nearly edge-on binary orientations to produce sufficiently close angular separations.  In this paper, we explore the properties and detectability of lensing signatures in simulated populations of MBHBs \citep{Kelley+2017a, Kelley+2019}, where we consider full distributions of binary parameters.
    
    Our models are based on the Illustris cosmological hydrodynamic simulations \citep{vogelsberger2014a} which provide self-consistently derived populations of MBHs \citep{sijacki2015} in merging galaxies along with their fully co-evolved stellar, gaseous, and dark matter content.  The MBH from merging galaxies are evolving in post-processing using sophisticated models of binary evolution based on the radial phase-space distributions of material in post-merger galaxies.  Our binary models include dynamical friction, loss-cone stellar scattering (with parameterized refilling), circumbinary disk torques based on evolving MBH accretion rates, and gravitational wave emission.

    A large number of uncertainties remain, particularly governing the dynamics of the accretion disks surrounding MBH binaries once they reach sub-parsec scales.  A general consensus model is beginning to take shape \citep{Paczynski-1977, Artymowicz+Lubow-1994, Artymowicz+Lubow-1996, Gould+Rix-2000, Armitage+Natarajan-2002, Hayasaki+2007, Roedig+2011, Farris+2014, Duffell+2014, DOrazio+2016, Munoz+2019}, though the results of modern, high-resolution and long-duration simulations are finding results significantly different from those of earlier studies.  The emerging picture, based largely on the results of 2D idealized hydrodynamic simulations, is that of a circumbinary accretion disk in which a central low density cavity (`gap') is cleared by the presence of the orbiting binary.
    
    Around each MBH component, a circumsingle disk forms, fed by accretion streams entering the cavity from the circumbinary disk.  In close binaries, once the dynamical time and even binary evolution timescale are shorter than the viscous time, the accretion is dynamically driven and can continue throughout inspiral \citep{Farris+2015}.  Hydrodynamic simulations show that accretion tends to favor the lower-mass component, and vary periodically near a few times the orbital period.  The effects of varying eccentricity \citep[e.g.][]{Munoz+2019, Zrake+2021, DOrazio+Duffell-2021} and more complex geometries \citep[e.g.][]{Moody+2019} are beginning to be explored in simulations, but detailed thermal structures, realistic cooling and feedback effects, magnetic fields and self-consistently derived viscosity \citep[c.f.][]{ShiKrolik:2015, Noble+2021}, and other factors have yet to be explored on timescales that probe important secular processes.
    
    Even individual AGN are complex and characterized by large intrinsic time-variability \citep{Barr+Mushotzky-1986, Hook+1994, Ulrich+1997, Kelly+2009, MacLeod+2010}, and particularly by substantial variance across populations \citep[e.g.,~historically,][]{Seyfert-1943, Fanaroff+Riley-1974}.  In the `unified AGN' paradigm \citep{Antonucci-1993, Urry+Padovani-1995, Fossati+1998}, accretion rate and especially viewing angle are believed to be the primary drivers of phenomenological differences.  It is generally believed that accretion flows are only able to effectively cool, become geometrically thin, optically thick, and radiatively efficient for $10^{-2} \lesssim \fedd \equiv \dot{M} / \dot{M}_\trt{Edd} \lesssim 1$.  These systems thus dominate most observational (particularly near-optical) populations.  Within the thin disk, \citet{Shakura+Suyaev-1973} (and similar) models, the temperature of radiating gas decreases farther away from the MBH, leading to stratified emission with bluer bands emitted closer in.  The optical emission from AGN is observed to vary significantly in brightness, over timescales as short as the light-crossing time of the disk at small radii.  At large radii, outside of a characteristic sublimation radius (typically $10^{-2} \lesssim r/\pc \lesssim 10$), dust forms and the accretion flow becomes geometrically thick \citep{Barvainis-1987, Suganuma+2006, Netzer-2015}.  This `dusty torus' obscures the bright inner disk for viewing angles near the disk plane.
    
    Much of the detailed structure and dynamics of AGN accretion flows and emission are only broadly agreed upon qualitatively, with substantial variations existing between different models.  In this analysis, we combine detailed models of MBH binary populations (Sec.~\ref{sec:meth_mbhbs}) with analytic Shakura-Sunyaev disks (Sec.~\ref{sec:meth_disk}), and models for lensing (Sec.~\ref{sec:meth_lens}) and Doppler variability (Sec.~\ref{sec:meth_dopl}).  To assess detectability, we adopt a largely phenomenological approach based on observed AGN variability (Sec.~\ref{sec:meth_drw}) and typical survey parameters (Sec.~\ref{sec:meth_det}).  In our results, we highlight expected detection rates, observable populations properties, and the plausible characteristics of observable lensing signatures.  In our discussion, we focus on the key assumptions and uncertainties included in our models, and also some of the important systematics external to our analysis, and speculate on some possible use cases of eventual self-lensing detections.


\section{Methods}

    \subsection{MBH Binary Populations}
        \label{sec:meth_mbhbs}
    
        We use populations of MBHBs derived from the Illustris cosmological hydrodynamical simulations which self-consistently co-evolve thousands of galaxies and their central black holes in a fixed comoving volume of $\lrs{106.5 \, \rm{Mpc}}^3$ \citep{vogelsberger2014b, genel2014, torrey2014}\footnote{The Illustris data, including MBH populations, are publicly available online: \href{https://www.illustris-project.org}{www.illustris-project.org} \citep{nelson2015}.}.  Massive black holes `seeds' of mass $1.4\E{5} \, \msol$ are placed in the center of dark-matter halos once their mass exceeds $7\E{10} \, \msol$.  The black holes then grow through accretion, calculated using a local Bondi-rate estimator, and through mergers with other MBHs.  Black hole particles are artificially fixed to the centers (potential minima) of their host galaxies to avoid spurious scattering from star and dark-matter particles.  The resulting MBH population, described in detail in \citet{sijacki2015}, has properties matching local MBH--galaxy scaling relations (such as $M$--$M_\trt{bulge}$), the inferred local mass-density of MBHs, and produce accretion luminosities consistent with the observed quasar luminosity function.
        
        In the Illustris simulations, once two MBH particles come within a gravitational softening length of one another (typically $\sim 0.1 - 10 \, \rm{Kpc}$), they are combined into a single MBH remnant.  Physically, the binary merger process from $\sim\!10^3 \, \rm{pc}$ down to actual coalescence at $\lesssim\!10^{-3} \, \rm{pc}$ involves complex dynamical interactions with the nuclear environments of the post-merger host galaxy \citep{Begelman+1980}.  We incorporate these dynamics using semi-analytic binary evolution models developed in \citet{Kelley+2017a, Kelley+2017b, Kelley+2019}.  Each MBH `merger' is identified in the output of the Illustris simulations, and its binary evolution is modeled in post-processing.  In addition to the basic MBH parameters (the masses, redshift and separation at the time of numerical-merger), the radial density and velocity profiles are calculated for the gas, stars and dark matter in the post-merger host galaxy.  These profiles are then used to calculate binary `hardening' rates ($da/dt$) due to dynamical friction \citep{Chandrasekhar1943, Binney+Tremaine-1987}, `loss-cone' stellar scattering \citep{Magorrian+Tremaine-1999, Merritt2013}, circumbinary disk torques \citep{Gould+Rix-2000, Haiman+2009}, and gravitational wave emission \citep{Peters-1964}.  Note that throughout this analysis we restrict ourselves to circular orbits.
        
        The results of this modeling are binary evolution histories for $\sim 10^4$ binaries with total masses between $\approx 10^6$--$10^{10} \, \msol$.  These binaries represent those in a finite volume, evolving over all redshifts.  From this sample, the total number of binaries in the observer's light-cone can be calculated \citep[see Sec.~2.3 of][]{Kelley+2019}.  To construct a full population, we use kernel density estimation to `resample with variation' for an arbitrary number of stochastic realizations of the universe \citep{Kelley-2021}.  The parameters of the full population\footnote{One realization of which is available online, \href{https://doi.org/10.5281/zenodo.4068485}{zenodo.4068485} \citep{Kelley-2020_kde-data}} are shown in Fig.~\ref{fig:pop}.
        
        We review some of the key features of the binary population here, but see \citet{Kelley+2017a} for a detailed description.  The total-mass and mass-ratio of binary systems are strongly anti-correlated.  This is partially due to selection effects---because we only consider MBH components with $m_i > 10^6 \, \msol$, extreme mass ratios can only occur in the most massive systems---and partially cosmological as very massive elliptical galaxies tend to experience a large number of minor/extreme mergers with satellites \citep{Rodriguez-Gomez+2015}.  To a degree, this is counter balanced by the tendency of extreme mass-ratio mergers to \textit{stall}, largely in the $\sim\mathrm{Kpc}$ regime where dynamical friction can become ineffective, especially due to tidal-stripping of the secondary galaxy \citep[see also,][]{Tremmel+2018}.  Overall, binaries that are both more massive and nearer equal mass are more effective at reaching the bound binary stage ($\lesssim 10 \, \mathrm{pc}$), and eventually coalescing.  As binaries harden from parsec-scales to milliparsecs (orbital periods of $\sim \mathrm{yrs}$), the `residence time' ($\tau = a / (da/dt)$) rapidly decreases, meaning that the expected number of systems rapidly declines at smaller separations\footnote{In the GW driven regime, for example, $N\propto \tau \propto a^4$.}.  Thus the vast majority of binary systems will always tend to be at the largest orbital periods being considered.
    
    \subsection{Accretion Disk Model}
        \label{sec:meth_disk}
    
        As in \citet{dorazio+distefano-2018} and \citet{Kelley+2019} we use the standard \citet{Shakura+Suyaev-1973} thin-disk emission model, where the temperature of the disk as of function of radius $r$ is calculated as,
        \begin{equation}
            \sigma T^4 = \frac{3Gm_i\,\dot{m}_i}{8\pi r^3} \lrs{1 - \scale[1/2]{r_{\trt{ISCO},i}}{r}}.
        \end{equation}
        Here $\sigma$ is the Stefan-Boltzmann constant, $\dot{m}_i$ is the accretion rate in the disk around an MBH with mass $m_i$, $r_{\trt{ISCO},i}\equiv 3 r_{s,i} = 6 Gm_i / c^2$ is the inner-most stable circular orbit (ISCO; assuming the spin of the MBH is zero).  The rest-frame spectral luminosity, in the presence of a magnification field $\mathcal{M}$, is then found by integrating the Planck function $B_\nu(T)$,
        \begin{equation}
            \label{eq:spec-lum}
            L_\nu = \pi \sinpar{\incdisk} \int_0^{2\pi} \int_0^\infty B_\nu(T) \, \mathcal{M}(r,\phi) \, r \, dr \, d\phi.
        \end{equation}
        In the absence of lensing, $\mathcal{M} = 1$.
        
        Here the inclination of the disk relative to the observer is $\incdisk$, with $\incdisk=\pi/2$ corresponding to a face-on disk.  Basic angular momentum arguments suggest that the inclination of the secondary AGN disk should be \textbf{aligned} with that of the binary orbit, i.e.~$\incdisk \approx \incorb$.  In dedicated circumbinary disk simulations this is more often an assumption than a result, and indeed models that consider misaligned feeding of the circumbinary disk from large scales, and higher order precession and non-coplanar torques \citep[\eg,~Lense-Thirring precession][]{BardeenPetterson:1975} suggest that \textbf{misaligned} inclinations (i.e.~$\incdisk \neq \incorb$) are possible \citep[\eg,][]{Nixon_tearSngl+2012, Nixon_tearCBD+2013}.  In our analysis we explore the implications of both aligned and misaligned geometries, which do end up producing substantial differences in model predictions.

        We assume there is no disk within the ISCO, and the outer edge of the disk is set by the Hill radius \citep{Eggleton1983},
        \begin{equation}
            R_{\trt{Hill},i} = a \, \frac{0.49 \, q_i^{2/3}}{0.6 \, q_i^{2/3} + \ln\!\lr{1 + q_i^{1/3}}}.
        \end{equation}
        Here $a$ is the semi-major axis, and $q_i$ is the ratio of object $i$'s mass to that of the other (e.g.~$q=q_2=m_2/m_1$).
        
        Our MBH population from Illustris includes accretion rates derived from the local galaxy gas properties at the resolution scale of the simulations ($\sim \mathrm{Kpc}$).  While our modeling continues to evolve MBH binaries in post-processing at smaller separations, recall that the Illustris simulations combine the two MBHs into a remnant.  Thus, after this time, we do not have accretion rates onto each MBH component, but instead onto the combined remnant.  We calculate accretion rates onto each component based on the results of detailed 2D hydrodynamic simulations.
        
        We use (1) an analytic function calculated in \citet[][Eq.~1]{Kelley+2019} that is motivated by the accretion ratio ($\lambda \equiv \dot{M}_2 / \dot{M}_1$) found in \citet{Farris+2014}, and (2) a fit to the accretion ratio found in \citet{Duffell+2020}\footnote{Also consistent with \citet{Munoz+2020}, see the comparison in Fig.~1 of \citet{Siwek+2020}.}.  The Farris~et~al.~results suggest a \textbf{peaked} accretion ratio (near $q\sim0.1$) that then decreases again at smaller mass ratios.  \citet{Duffell+2020} represents an updated and more sophisticated calculation using the same code, and give results that are \textbf{monotonic} with mass ratio: \mbox{$\lambda \approx \lr{0.1 + 0.9 q}^{-1}$}, based on simulations with \mbox{$q \gtrsim 0.02$}.  Theoretical expectations, however, and the results of planet-star studies with $q \ll 1$, suggest that in the limit of \mbox{$q\rightarrow0$} the ratio \mbox{$\lambda \rightarrow 0$} \citep[e.g.][]{Kley+2003}. For this reason, we continue to consider the `peaked' model which includes a power-law decline at small $q$.  Note that the exact location and shape of the turnover is highly uncertain, and might not be well captured by the \citet{Kelley+2019} fitting function.

        The Illustris accretion rates are limited to Eddington.  In our fiducial model, we do not enforce this criteria on each component MBH, and thus as $\lambda$ is often greater than one, the secondary is allowed to be super-Eddington.  We use the exact same treatment of the disk in the super-Eddington case as in sub-Eddington accretion.

    \subsection{Gravitational Lensing}
        \label{sec:meth_lens}
    
        To calculate lensing lightcurves, we use the gravitational lensing models described in \citet{dorazio+distefano-2018}.  In the point-source and weak-field limit, the lensing magnitude is,
        \begin{equation}
            \label{eq:mag}
            \mathcal{M} = \frac{u^2 + 2}{u\,\lr{u^2+4}^{1/2}},
        \end{equation}
        for a projected separation $u = \rm{Re}\lr{u_1 + u_2}$, in units of the Einstein radius.  The (complex) projected separation of each component of the binary is,
        \begin{equation}
            u_i = \lrs{\frac{a \, c^2 \lr{\cos^2\phi_i + \sin^2\incorb\cos^2\phi_i} }{4G(M-m_i) \cos \incorb \sin\phi_i} }^{1/2},
        \end{equation}
        for an orbital inclination relative to the line-of-sight $\incorb$ (such that $\incorb=0$ is an edge-on orbit), orbital phase of each component $\phi_i$ (thus $\phi_1 \equiv \phi_2 \pm \pi$), speed of light $c$, and gravitational constant $G$.  The combined binary mass $M = m_1 + m_2$.  We define the phase such that $\phi_i = \pi / 2$ corresponds to component $i$ being directly between the observer and its companion.  The Einstein radius of each component is,
        \begin{equation}
            \label{eq:erad}
            R_{E,i} = \lr{2 a \, r_{s,i} \cos\incorb \sin\phi_i}^{1/2}.
        \end{equation}
        Finally, the duration of the lensing event can be approximated as,
        \begin{equation}
            \label{eq:lens_dur}
            \tlensi = \frac{\pobs}{\pi} \sin^{-1}\scale{R_{E,j}}{a},
        \end{equation}
        where the subscripts $i$ and $j$ are used to indicate that the duration of the lensing event for one component depends on the Einstein radius of the other.
        
    \subsection{Doppler Variability}
        \label{sec:meth_dopl}

        The Doppler boosted flux can be calculated as \citep{DOrazio+2015},
        \begin{equation}
        F'_{\nu'} = D^3 F_\nu,
        \end{equation}
        where the observed frequency $\nu'$ is related to the rest-frame frequency as $\nu' = D \nu$.  The Doppler factor is defined as \mbox{$D = \lrs{\gamma \lr{1 - v_o/c}}^{1/2}$}, the Lorentz factor \mbox{$\gamma \equiv \lr{1 - v^2 / c^2}^{-1/2}$}, and the line of sight velocity \mbox{$v_o = v \cos{\incorb} \cos{\phi}$}.  For all of our quantitative calculations, we calculate the Doppler boosted flux using the spectrum from Eq.~\ref{eq:spec-lum} at the appropriately shifted frequencies.  This is well approximated by the relation \citep{Charisi+2018},
        \begin{equation}
            \frac{F'_{\nu'}}{F_\nu} \approx \lr{3 - \alpha_\nu} \frac{v_o}{c} + 1,
        \end{equation}
        where $\alpha_\nu$ is the spectral index at the frequency of interest.  For frequencies corresponding to the typical black-body temperatures in the disk, $\alpha_\nu \approx 1/3$, which is the spectral slope at the peak emission frequencies for a steady-state, optically thick, geometrically thin accretion disk  \citep[\eg,][]{FKR_APIA:2002}.  The optical spectral index measured in observed AGN is seen to vary significantly \citep[e.g.,][]{Charisi+2018}, but to be broadly consistent with $1/3$ which we adopt throughout our analysis.

    \subsection{AGN Intrinsic Variability (Damped Random Walk)}
        \label{sec:meth_drw}
    
        We model the intrinsic variability of AGN as a damped random walk following \citet{MacLeod+2010}.   In particular, the probability distribution for the lightcurve magnitude a time $\Delta t$ after a previous value at time $t$ is characterized by an expectation value and variance given by (\textit{Ibid.}~Eq.~5),
        \begin{equation}
            \begin{split}
                E\lrs{ M_\nu(t + \Delta t) \, | \, M_\nu(t) } & = M_\nu(t) \, e^{-\Delta t / \tdrw} + \mean{M_\nu} \lr{1 - e^{-\Delta t / \tdrw}} \\
                \mathrm{Var}\lrs{ M_\nu(t + \Delta t) \, | \, M_\nu(t) } &= \frac{1}{2} \lr{\sfinf}^2 \, \lr{1 - e^{-2 \Delta t / \tdrw}}.
            \end{split}
        \end{equation}
        Here the magnitude $M_\nu(t) \propto \logten{L_\nu\lrs{t}}$, and has a mean value of $\mean{M_\nu}$, $\tdrw$ is the characteristic correlation time-scale, and $\sfinf$ is the structure function\footnote{A measure of self-similarity used analogously to an autocorrelation function.} as $\Delta t \rightarrow \infty$.
        In the limit of small and large time spans, the standard deviation becomes (\textit{Ibid.}~Eq.~4)\footnote{Note that `$\sigma$' in Eq.~4 of \citet{MacLeod+2010} refers specifically to what we are calling $\sigmadrw\lr{\Delta t \gg \tdrw}$},
        \begin{equation}
            \label{eq:drw_noise}
            \begin{split}
                \sigmadrw\lr{\Delta t \ll \tdrw} & = \sfinf \, \scale[1/2]{\Delta t}{\tdrw}, \\
                \sigmadrw\lr{\Delta t \gg \tdrw} & = \sfinf \, 2^{-1/2}.
            \end{split}
        \end{equation}
        Thus the power spectrum decreases for variations shorter than $\tdrw$ (i.e.~`red' at higher variation-frequencies), but approaches a constant value for longer timescales (i.e.~`white' at lower frequencies).  In this context, the most important feature of red-noise spectra is their natural tendency to produce coherent excursions from the mean at longer timescales.  These excursions can easily resemble order unity cycles of sinusoidal variations (like Doppler boosting) or one-off peaks (like lensing flares).  In red-noise regimes of parameter space, multiple complete cycles, repeated flares, or external constraints are often needed to confirm signals.

        \citet{MacLeod+2010} provide empirical scaling relations for the damped random walk parameters as a function of wavelength, brightness, and MBH mass.  To use these relations, we calculate i-band magnitudes with the bolometric corrections from \citep{Runnoe+2012}.  The distribution of DRW parameters calculated from our binary population are shown in Fig.~\ref{fig:app_drw} for reference.  The interquartile range of $\tdrw$ is $[31, 68] \, \mathrm{days}$ and $[0.13, 0.32] \, \mathrm{mag}$ for $\sfinf$ (corresponding to $[24\%, 60\%]$ in flux). 

    \subsection{Detection Criteria}
        \label{sec:meth_det}
        
        The Illustris simulations and subsequent binary evolution yields masses $m_1$, $m_2$, redshift $z$, observed orbital period $\pobs$, Eddington factor $\fedd$, and orbital inclinations are randomly distributed over all solid angles.  We also assume circular orbits throughout.  We use either the `monotonic' or `peaked' accretion ratio functions to determine the secondary accretion rate ($\feddsec$), and either orient all AGN disks to be `aligned' with the binary orbit, or alternatively to be independently, randomly-distributed over all angles in the `misaligned' configuration.
        
        We determine detectable systems as those matching the following criteria.  The observed orbital period must be less than $\pobsmax$, which is $5 \, \mathrm{yr}$ in our fiducial model.  Longer orbital periods are considered later (Sec.~\ref{sec:long}), but in general we assume that multiple lensing flares will be needed to confirm the signal type, and thus orbital periods will need to be shorter than half the total survey duration.  Our fiducial value of five years is motivated by the planned $10 \, \mathrm{yr}$ duration of VRO's LSST \citep{Ivezic+2019_lsst}, and the roughly decade of observations of recent all-sky surveys such as CRTS and PTF.  The total flux of the secondary accretion disk (including truncation at the Hill radius) must be above the target survey sensitivity $\fsens$.  Additionally, the change in observed flux must exceed a threshold,
        \begin{equation}
            \label{eq:snr}
            \begin{split}
                \frac{\Delta F_\nu}{F_\nu} & > \frac{1}{\snr} + 0.05, \\
                \snr & \equiv \frac{F_\nu}{\fsens + \fnoise}, \\
                F_\textrm{noise} & = F_\nu \cdot \sigmadrw\lr{\Delta t}.
            \end{split}
        \end{equation}
        The higher the SNR, the smaller the detectable flux variations \citep[e.g.][]{Sarajedini+2003, Liu+2016}, but we assume a maximum sensitivity to variations of $5\%$ \citep[e.g.][]{Graham+2015}.  Note that due to the nature of DRW noise, the amplitude is dependent on the timescales of interest.  For lensing, $\Delta t = \tlens$, while for Doppler variations, $\Delta t = \pobs$.  In the latter case, generally $\pobs / \tdrw \gg 1$, and the DRW noise approaches a nearly constant value (Eq.~\ref{eq:drw_noise}).

    \begin{figure*}
        \centering
        \includegraphics[width=1\textwidth]{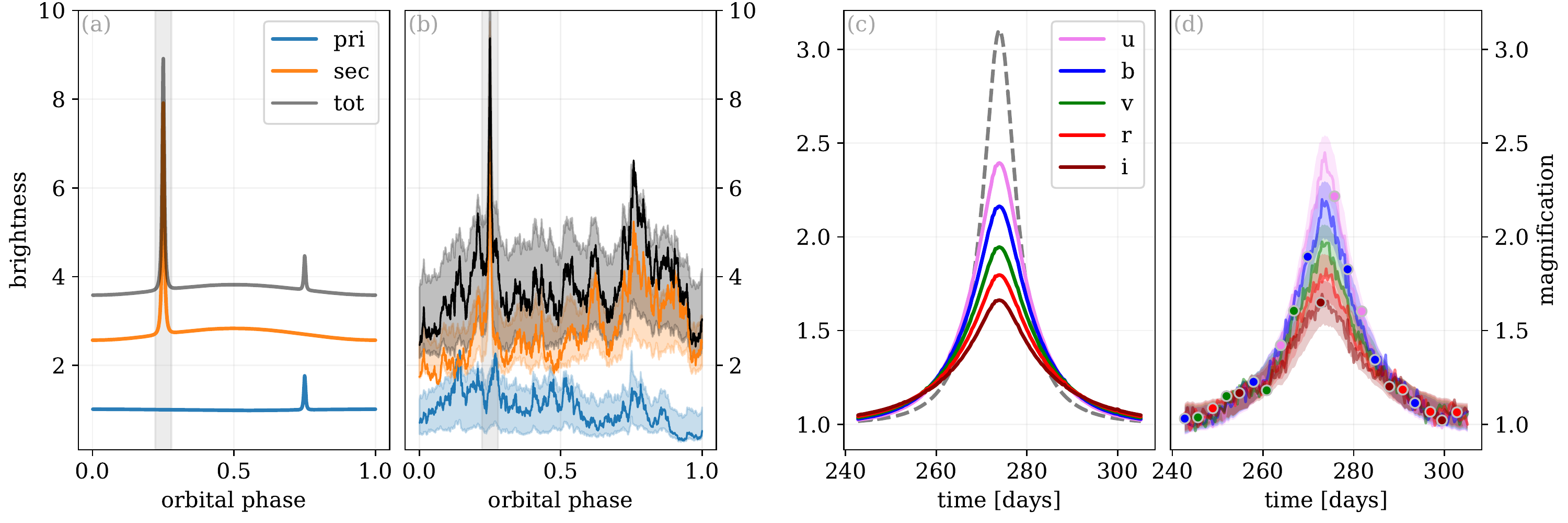}
        \caption{\textbf{Lensing lightcurve for an example MBH binary}: luminosity in arbitrary units versus time.
        \textbf{(a)} Point-source approximation for the \textcolor{mplblue}{primary (blue)}, \textcolor{mplorange}{secondary (orange)}, and total luminosity (black), including both lensing and Doppler boosting.  The first, larger lensing peak corresponds to the secondary lensed by the primary.  \textbf{(b)} Point-source lightcurves with added damped-random-walk (DRW) variations, showing a random realization (line) and the two-sigma range (shaded band).  \textbf{(c)} Lightcurves by SDSS band using a finite-size, thin-disk emission model.  The dashed black curve is the point-source approximation. \textbf{(d)} Finite-size light curves including DRW variations, with the shaded bands indicating the inter-quartile range.  The solid lines show a single DRW realization applied to each band's lightcurve, with an additional independent $2\%$ random noise.  The scatter points show time-sampling with 3-day cadence, randomly chosen across the different bands.  Binary parameters: $M=10^8 \msol$, $q=0.3$, $\pobs=3\yr$, $z=0.2$, $\incorb=0.02$, $\incdisk=0.4\pi$, $\fedd=0.25 \,\, (\feddsec=0.79)$.  The projected closest approach of the secondary is $0.6 R_{E,\textrm{prim}}$.}
        \label{fig:lightcurve}
    \end{figure*}


\section{Results}

    \subsection{Typical Lensing Features}
    
        An example AGN lightcurve including lensing and Doppler boosting is shown in Fig.~\ref{fig:lightcurve}.  Panel (a) shows the point-source approximation for a bolometric lightcurve.  The first, larger lensing peak ($\phi/2\pi = 0.25$) is produced by the primary lensing the secondary, after which the secondary begins moving towards the observer as the Doppler boosting signature increases towards a maximum ($\phi/2\pi = 0.5$).  In this case the peak magnification in the point-source approximation is $\mathcal{M}_2 = 3.10$ for the secondary (lensed by the primary), and $\mathcal{M}_1 = 1.85$ for the primary.  The  In this example we choose an overall Eddington fraction for the binary, $\fedd = 0.25$, and calculate the relative accretion rates to the primary and secondary following the `peaked' model, giving $\feddsec(q=0.3) = 0.79$.  The total brightness increase at the secondary's peak is $2.41$ at a closest approach of $0.3$ Einstein radii, and that of the primary's is $1.21$ at $0.6$ Einstein radii.

        The AGN lightcurve with the addition of typical DRW noise is shown in Fig.~\ref{fig:lightcurve}(b).  Applying the DRW scaling-relations from \citet{MacLeod+2010} to the secondary MBH\footnote{The full distribution of DRW parameters for each simulated binary's secondary MBH are plotted in Fig.~\ref{fig:app_drw}.}, the characteristic damping timescale for this system is $\tdrw = 77 \, \textrm{days}$ with a standard deviation at large timescales of $\sigmadrw = \sfinf / \sqrt{2} = 0.12$ magnitudes, or $\approx 22\%$ in sensical measures of brightness.  The first lensing peak is clearly distinguishable above the noise, while the second (the lensing of the primary by the secondary) is completely washed out.  Note that for very small angular separations, the ratio of peak lensing magnitudes $\mathcal{M}_2 / \mathcal{M}_1 \approx q^{-1/2}$, and further that the accretion partition functions tend to have the secondary significantly brighter than the primary.  For these reasons, the remainder of our analysis considers the detectability of only the secondary MBH's accretion disk being lensed by the primary.

        Panels (a)~\&~(b) of Fig.~\ref{fig:lightcurve} both use the point-source approximation applied to bolometric luminosities.  Panels (c)~\&~(d) include finite-size effects and compare the brightness between different SDSS bands.  Panel (c) is idealized while panel (d) includes DRW noise using the same parameters as in (b), but now applied to each band. The analytic point-source magnification for the secondary is $3.1$, but when considering the finite-size leads to magnifications between $1.66$ in the i-band and $2.39$ in the u-band.  In general, bluer bands are magnified more than redder bands, even when the projected closest-approach of the lens occurs well outside of where most of the bluer emission is being produced.  While the highest magnification may occur at larger (redder) radii, the typical magnification will be higher in the blue bands which are produced in more compact regions.  The color effects across the population are discussed further below.
        
        In panel (d) of Fig.~\ref{fig:lightcurve}, an example multi-band time-sampling is also marked with dots.  `Observations' are chosen in a single random band, at an interval of every 3 days.  The overall cadence is consistent with preliminary expectations for the LSST survey, but the detailed survey parameters are yet to be determined.  For reference the analytic duration for this event (Eq.~\ref{eq:lens_dur}) is $11$ days, while the point-source full-width at half-magnitude is $8.1$ days, and that of the finite-size curves (neglecting noise) ranges from 13 days in the u-band to 18 days in the i-band.

    \begin{figure*}
        \centering
        \includegraphics[width=\textwidth]{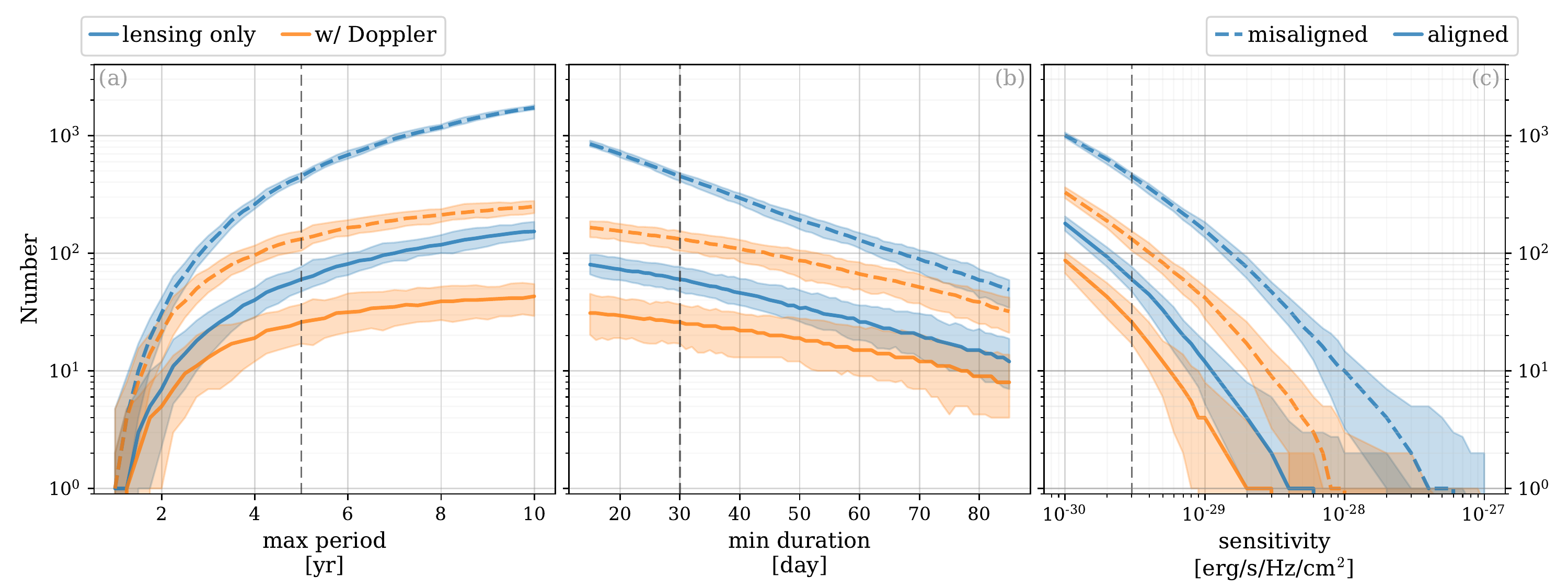}
        \caption{\textbf{Detectable systems for varying survey sensitivities}: the number of binaries with \textcolor{mplblue}{detectable lensing signatures (blue)} and \textcolor{mplorange}{detectable lensing and Doppler variations (orange)}.  Lines show median values while shaded regions are $2-\sigma$ contours.  Each panel varies one parameter while keeping the other two fixed, where the fiducial values are: observed orbital periods $p_\trt{obs} < 5$ yr, lensing signal durations $\tau > 50$ days, and R-band fluxes $F_\nu > 3.0\E{-30} \, \fluxunits{}$.  When secondary MBH circum-single disks are perfectly aligned with the binary orbital plane (solid lines), sources are up to an order of magnitude less common than if the secondary disks are randomly oriented (dashed).}
        \label{fig:surveys}
    \end{figure*}
    
    \subsection{Detection Rates and Populations}
    
        Binary self-lensing detection rates are shown as a function of survey parameters in Fig.~\ref{fig:surveys}.  We show the all-sky number of detectable sources both for lensing signals alone (blue) and cases where both the lensing and Doppler signals are detectable (orange).  Generally, we assume that a survey will need to see multiple, periodic lensing flares to confirm that it is not a flare due to intrinsic variability, or a transient (i.e.~one-off) lensing event.  For high magnification and long-duration events, this will be a conservative assumption.  For a survey with a $10\yr$ duration, requiring two events implies a maximum orbital period of $5\yr$, which we adopt for our fiducial parameters.  \textbf{Typical all-sky detection rates for LSST sensitivity are \vpm{60}{+17}{-13} (\vpm{450}{+29}{-40}) lensing, and \vpm{26}{+11}{-9} (\vpm{130}{+22}{-27}) with Doppler, for the aligned (misaligned) configuration}. Note that LSST is expected to cover roughly $1/4$ of the sky \citep{Ivezic+2019_lsst}.  The included uncertainties are at the $2-\sigma$ level based on detections from 100 realizations of our binary populations, which accounts for Poisson-like variations in the distributions of binary properties, but does not account for uncertainties in model parameters.
        
        Because the binary hardening timescale, and thus the number of binaries at a given orbital period, depends strongly on the orbital period ($N \propto \pobs^{8/3}$), the resulting detection rate is quite sensitive to the maximum detectable orbital period.  If the fiducial maximum orbital period decreases by a factor of two (\mbox{$\pobs < 2.5 \yr$}), the all-sky numbers drop to \vpm{14}{+8}{-7} (\vpm{66}{+17}{-15}) lensing, and \vpm{9.5}{+6.5}{-5.5} (\vpm{39}{+13}{-11}) with Doppler.  The lensing population more closely follows the overall period-distribution of binaries than the Doppler population, because the Doppler detectability increases for smaller (faster) orbits.
        
        Lensing flare durations and the DRW damping timescales are both typically 30--100 days in duration.  However, including DRW noise only has a moderate effect on lensing detection rates.  Setting $\fnoise = 0$ in Eq.~\ref{eq:snr} (compare `white' vs. `DRW' in Table~\ref{tab:tab}) increases the detection rate from a fiducial value of \vpm{60}{+17}{-13} (\vpm{450}{+29}{-40}) to \vpm{73}{+16}{-13} (\vpm{510}{+31}{-44}) for the aligned (misaligned) configuration.  The change in Doppler (alone) detection rates, however, is almost a factor of two, from \vpm{160}{+23}{-20} (\vpm{370}{+36}{-32}) to  \vpm{220}{+31}{-18} (\vpm{670}{+51}{-38}).  The difference is that Doppler boosts typically have magnitudes on the order of 10s of percent while lensing magnifications are typically order unity.  The effect on the combined detection rate changes from \vpm{26}{+11}{-9} (\vpm{130}{+22}{-27}) to \vpm{44}{+15}{-10} (\vpm{250}{+32}{-26}).  Lensing signatures may be more robust against DRW noise compared to Doppler variability, but this would need to be confirmed using a realistic detection pipeline.

        We assume that some minimum number of detections spanning the lensing event are required to properly characterize it.  In Fig.~\ref{fig:surveys}b, we parameterize this criterion in terms of the minimum lensing event duration that the survey is sensitive to. In practice, this is the number of intra-flare photometric measurements required for detection, multiplied by the survey cadence.  As a fiducial value, we assume 10 intra-flare measurements are required with a cadence of 3 days \citep[roughly that expected for LSST;][]{Ivezic+2019_lsst}, yielding a minimum duration of 30 days.  If this value is halved to only 15 days, the all-sky number of events is \vpm{80}{+17}{-14} (\vpm{850}{+58}{-43}) lensing, and \vpm{31}{+14}{-11} (\vpm{160}{+22}{-29}) with Doppler.  Fig.~\ref{fig:surveys}c shows the detection rate dependence on r-band survey sensitivity.  The fiducial value of $3\E{-30} \fluxunits$ is motivated by the expectation for LSST.  If this can be boosted to $2\E{-30} \fluxunits$ detections increase proportionally to \vpm{93}{+19}{-16} (\vpm{630}{+42}{-58}) lensing, \vpm{42}{+13}{-15} (\vpm{190}{+29}{-24}) with Doppler.  An order of magnitude decrease in sensitivity, however, comparable to the sensitivity of SDSS, yields all-sky numbers of \vpm{2.0}{+4.0}{-2.0} (\vpm{47}{+10}{-12}) lensing, \vpm{1.0}{+1.0}{-1.0} (\vpm{9.0}{+6.0}{-5.7}) with Doppler.  Unless most secondary AGN disks are misaligned with their binary orbits, SDSS-like sensitivities are thus unlikely to detect MBHB lensing signatures.
    
        The parameters of simulated MBH binaries with detectable lensing signatures are plotted in Fig.~\ref{fig:pop} for both the `aligned' (green) and `misaligned' (orange) configurations.  In the aligned case, the median and interquartile range of lensing detectable systems are: $M = 1.1^{+1.3}_{-0.6}\E{9} \msol$, $q = 1.2^{+5.6}_{-0.8}\E{-2}$, $\pobs = 3.4^{+0.8}_{-0.9} \yr$, and $z = 0.46^{+0.24}_{-0.20}$.  Lensing detectable systems pick out higher masses and more extreme mass-ratios than the overall population.  For a given orbital separation, higher masses and more extreme mass-ratios lead to higher lensing probabilities as a larger range of inclination angles will produce equivalent magnifications\footnote{Note that Doppler signatures are also more detectable in the same population of systems due to higher orbital velocities \citep[see Fig.~10 of][]{Kelley+2019}.}.  The trend is further bolstered by larger masses and more extreme mass ratios producing more luminous secondaries (see, Sec.~\ref{sec:meth_disk}).  The correlation between Eddington factor and redshift, and anti-correlation with mass are produced by the same selection effect. In the misaligned geometry, the median and interquartile values are: $M = \mvpm{4.4}{+7.5}{-2.0}\E{8} \msol$, $q = \mvpm{3.5}{+7.9}{-2.7}\E{-2}$, $\pobs = \mvpm{3.7}{+0.6}{-0.8} \yr$, and $z = \mvpm{0.75}{+0.44}{-0.26}$.  The changes relative to the aligned case, particularly lower masses, larger (more-moderate) mass-ratios, and higher redshifts are all driven by the misaligned disks presenting larger angular areas and thus being brighter than for aligned disks.

        Figure~\ref{fig:pop} also shows the observed signal properties for aligned and misaligned models.  For a lensing signal to be detectable, the binary orbit must be nearly edge-on ($\incorb \approx 0$).  In the aligned configuration, the secondary AGN disk orientation is by definition the same ($\incdisk \equiv \incorb$), while in the misaligned case it is oriented randomly.  In the aligned configuration, however, as $\incdisk \rightarrow 0$ the flux from the AGN decreases as $F_\nu \propto \incdisk$ as a smaller and smaller angular area of the disk remains visible.  This effect could be significantly enhanced if high optical-depth obscuring torii are present in MBHBs as they are in single AGN \citep[see, e.g.,][]{Ueda+2014}.  This important issue is discussed further in Sec.~\ref{sec:disc}.  The resulting effect is that in the aligned case, $\incorb$ tends to be larger than in the misaligned case: $\incorb = \mvpm{9.5}{+4.7}{-3.5}\E{-2} \approx 5^{\circ}$ versus $\incorb = \mvpm{4.8}{+3.5}{-2.4}\E{-2} \approx 3^{\circ}$.  This leads to a proportional change in typical lensing magnitudes: $1.9^{+0.5}_{-0.4}$ (aligned) versus $2.5^{+1.5}_{-0.7}$ (misaligned).  The lensing duration, calculated from Eq.~\ref{eq:lens_dur}, is only weakly dependent on inclination and thus remains very similar in both geometries: $56^{+24}_{-15}$ days (aligned) and $46^{+18}_{-9}$ days (misaligned).

    \subsection{Finite-Size Effects and Chromatic Lensing}

        \begin{figure}
            \centering
            \includegraphics[width=\columnwidth]{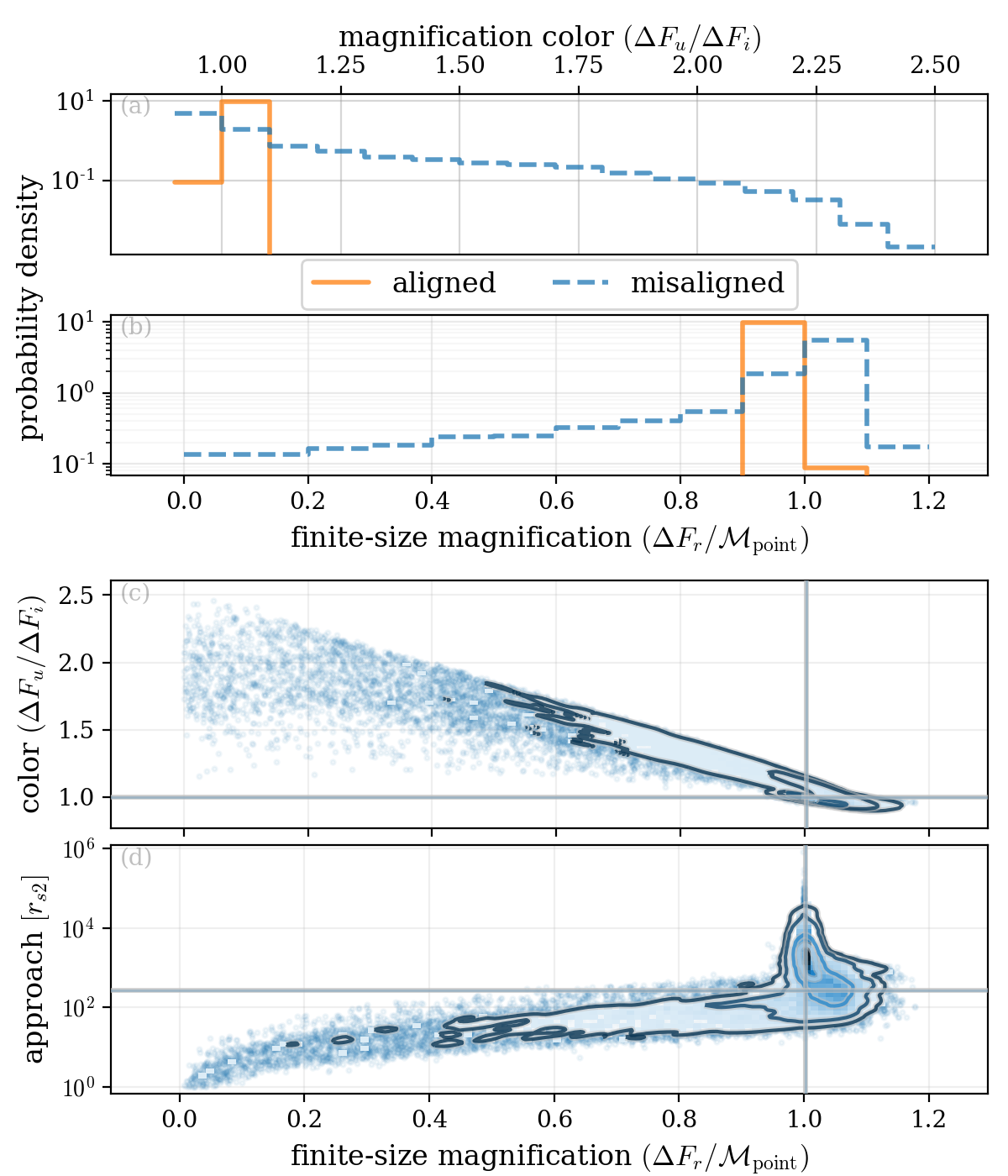}
            \caption{
            \textbf{Distribution of magnification colors, and magnification amplitudes relative to the point-source approximation.}  Panel (a) plots `color', defined as the ratio of magnifications in the u-band relative to the i-band.  Panel (b) shows the r-band magnification relative to the point-source approximation, and panel (c) shows the strong anti-correlation of this ratio versus magnification color.  Panel (d) plots closest approach (in units of secondary Schwarzschild radii) relative to finite-size magnification.  In panels (c) and (d), only misaligned systems are plotted because magnification and color are almost identically unity for aligned systems.  These distributions are shown for 100 realizations of detected systems, roughly 3000 (29,000) for aligned (misaligned) geometries.
            }
            \label{fig:magnif}
        \end{figure}

        The top panel (a) of Fig.~\ref{fig:magnif} shows the ratio of magnification in the u-band to that of the i-band.  Larger values correspond to AGNs appearing bluer at the lensing peak.  For the aligned configuration, lensing is almost perfectly achromatic: the u-band magnification is within $2\%$ that of the i-band in the interquartile range.  For aligned systems, the AGN disk is nearly edge-on, and the lensing magnification affects all disk radii (and thus colors) similarly.  For misaligned configurations, most systems are still nearly achromatic with the median and interquartile range: $\Delta F_u / \Delta F_i = 1.00^{+0.19}_{-0.02}$, but $10\%$ of systems have $\Delta F_u / \Delta F_i > 1.56$.  Half of lensing-detectable systems are redder at the lensing peak, though only by a small amount: $\Delta F_u / \Delta F_i = 0.98^{+0.01}_{-0.02}$.  This occurs in binaries where the peak emission radii of the redder bands are comparable to (i) the Hill radius of the secondary MBH, and also (ii) near the projected separation of the lens at closest approach.  In these cases, the secondary MBH's accretion disk is truncated, and the un-magnified emission is attenuated in the redder bands.  During lensing, the bluer bands are still brighter, but the relative increase is more substantial in the redder bands.
    
        Panel (b) of Fig.~\ref{fig:magnif} shows the disk magnification in the r-band compared to the magnification in the point-source approximation.  For aligned systems, the magnification is almost identical to the point-source approximation: in situations where detectable lensing occurs, the angular area of the nearly edge-on disk tends to be small, and different radii are projected to similar angular offsets.  For misaligned binaries, the majority of systems are clustered around the point source approximation: $1.00^{+0.03}_{-0.08}$.  Out of the misaligned systems, $58\%$ are magnified more than in the point-source approximation, but only slightly: the highest excesses reaching $F_r^\mathrm{lensed}/F_r^\mathrm{lensed,point} \approx 1.10$. These are the same binaries that result in lensing peaks that are reddened (as opposed to the more common bluened peaks). Both phenomena are caused by lensing of the disk just within the truncation radius which overcompensates for attenuation in the unlensed-case.  This can be seen in panel (c) of Fig.~\ref{fig:magnif}, which shows the strong anti-correlation between lensing color and magnification relative to the point-source approximation.

        To understand these reddened systems more carefully, consider the (projected) passage of the lensing primary across the plane of the secondary's accretion disk.  Take the projected closest approach of the primary to be just outside of the characteristic radius for r-band emission.  If the secondary disk extends to infinity (i.e.~it is not truncated), then the bluer emission comes from a more compact region which is lensed more strongly---even though the lens (primary) is closer to the r-band radius at one azimuth in the disk, the average projected distance to the bluer emitting regions is still smaller than the average distance to the r-band emitting regions.  For the infinite disk, there is a larger amount of r-band emitting material at larger radii (including on the opposite side of the disk) that is farther from the lens and magnified less.  If the secondary disk is instead truncated just beyond the closest approach radius (and the characteristic r-band radius), then there is no longer as much r-band emitting material at larger radii, and the relative magnification of the r-band emission is much higher.  Note, that these reddened systems are still magnified less than in the case of the point-source approximation applied to an un-truncated AGN disk.

        A noticeable tail in the population extends towards smaller magnifications with $10\%$ of systems having $F_r^\mathrm{lensed}/F_r^\mathrm{lensed,point} < 0.56$.  Smaller than point-source magnifications are caused by systems in which the minimum angular separation is very small, and the point-source approximation begins to diverge (Eq.~\ref{eq:mag}).  In actuality, only a small fraction of the AGN disk passes at these close separations to the lens, and only that fraction of the emission is magnified to high levels.  The correlation between closest projected approach distance at peak magnification and the finite-size magnification is shown in panel (d) of Fig.~\ref{fig:magnif}.  The point-source approximation is breaking down right when the closest approach is comparable and smaller than the peak emission radii of interest ($\sim 100 r_s$). See also the discussion of finite-sized effects in \citep{dorazio+distefano-2018}.

        An upper-bound to the wavelength-dependent maximum magnification can be calculated by setting the minimum angular separation to be the extent of the AGN disk at a given emitting wavelength, i.e.~$u_\mathrm{min}=\max\lrs{u_\textrm{orbit}, u_\textrm{disk}}$.  Using the radius of peak emission for a given frequency-band (here we've normalized to the SDSS r-band),
        \begin{equation}
            r_\textrm{peak}(\nu) = 16.0 \, r_{s,i} \; \feddi^{1/3} \, \scale[-1/3]{m_i}{10^8\,\msol} \scale[-4/3]{\nu}{4.6\E{14} \, \mathrm{Hz}},
        \end{equation}
        and approximating the magnification as $\mathcal{M} \approx 1 / u$,
        \begin{equation}
           \mathcal{M}_\mathrm{peak}^\mathrm{max} = 7.25 \, \scale[4/3]{\nu}{4.6\E{14} \, \mathrm{Hz}} \scale[1/3]{\pobs}{3\yr} \scale[-2/3]{q}{0.1} \frac{\lr{1 + q}^{1/6}}{\feddsec^{1/3}}.
        \end{equation}
        In practice the true peak magnification will be lower after convolving the full magnification field with the disk surface brightness. The maximum magnification at any wavelength can be estimated by considering the magnification of emission emanating from the ISCO of the source BH \citep{DoDi:2020},
        \begin{equation}
            \begin{split}
            \mathcal{M}^\mathrm{max} & = \frac{R_{E,1}}{3 \, r_{s,2}} \approx \frac{\lr{2 \, a \, r_{s,1}}^{1/2}}{3 \, r_{s,2}} \approx 7 \, \scale[1/2]{a}{100 \, R_s} \, \frac{\sqrt{1+q}/q}{\sqrt{2}}\\
                & \sim 100 \, \scale[1/3]{\pobs}{3 \, \yr} \scale[-1/3]{M}{10^8 \msol} \scale[-1]{q}{0.1} \lr{1 + q}^{1/2},
            \end{split}
            \label{Eq:Maxmag}
        \end{equation}
        where $R_s = 2GM/c^2$, and the top line assumes a conservative $q=1$.

        Note that these results will depend on accretion disk structure.  In this analysis we assume a Shakura-Sunyaev thermal structure, but observed lensing systems will be able to serve as probes of this assumption.
        
        \begin{figure}
            \centering
            \includegraphics[width=\columnwidth]{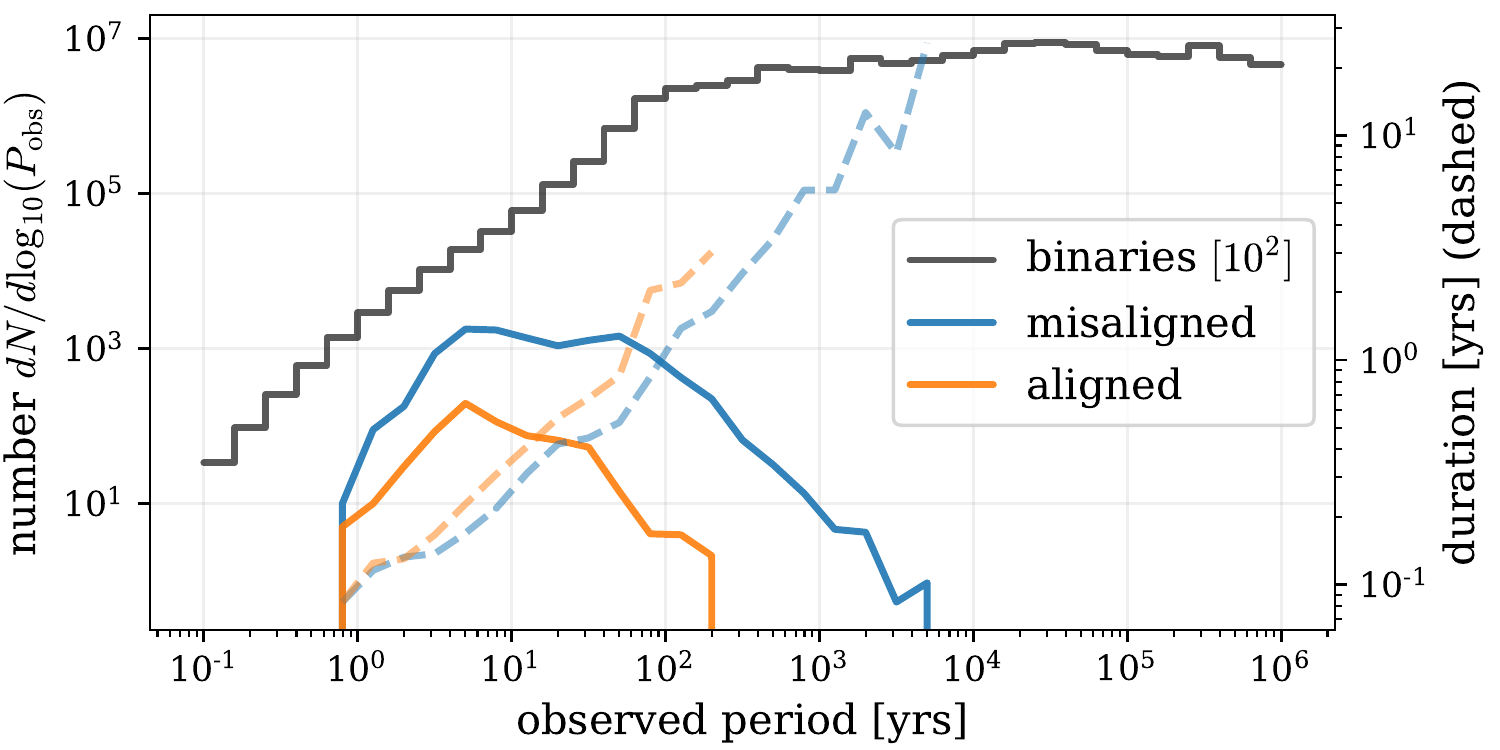}
            \caption{\textbf{Occurrence rate of long-period lensing binaries}.  Solid lines (left axis) show the number of all simulated binaries (black; shifted by $10^2$), and the number of lensing-detectable binaries in the \textcolor{mplblue}{misaligned (blue)} and \textcolor{mplorange}{aligned (orange)} configurations.  Dashed lines (right axis) show the average lensing duration at each observed orbital period.  We require lensing durations $\tlens > 30$ days, and assume an LSST-like sensitivity of $3\E{-30} \fluxunits$, but no longer impose a maximum orbital period (to ensure multiple detectable events).  We assume a $5 \yr$ observing duration both for calculating signal-to-noise (at short periods) and also the probability of the lensing event occurring during the observing window (at long periods).}
            \label{fig:long}
        \end{figure}
    
    \subsection{Long-Period, Long-Duration Signals}
        \label{sec:long}
    
        Our fiducial models and detection rates (i.e.~Table~\ref{tab:tab}) assume that multiple, periodically repeating flares must be identified to confirm a self-lensing binary origin.  
        One can imagine, however, that high signal-to-noise events and long-duration events where the lensing light-curve can be thoroughly sampled, could allow identification of a self-lensing binary with only one event. Such an event would then offer a prediction for the arrival of the next lensing flare \citep[\eg][]{HuSpikey+2020}.  Fig.~\ref{fig:long} shows the number of detectable lensing-binaries without requiring multiple events to be detectable.  The total binary population shows two regimes. At periods shorter than $\approx 10^2 \yr$, binaries are firmly in the gravitational wave driven regime and spend less and less time at smaller separations, and thus as a population there are ever fewer.  At longer orbital periods ($\gtrsim 10^2 \yr$), binary hardening rates are determined largely by interactions with their local environment, primarily in the form of stellar scattering and torques from a circumbinary disk, which end up producing more uniform hardening rates versus separation \citep[see, e.g.,][]{Kelley+2017a}.

        The distribution of lensing-detectable binaries cuts off sharply both at short and long orbital periods.  At short periods, the duration of lensing events becomes small, and even the high cadence of an LSST-like survey becomes unlikely to make enough observations during the lensing event to properly characterize it as such.  This effect impacts both aligned and misaligned systems in the same way.  At long periods, lensing signals are competing with the saturated levels of intrinsic red-noise (recall, modeled as a damped random walk; see Eq.~\ref{eq:drw_noise}).  
        At the same time, even though the maximum possible lensing magnification goes up for larger binary separations (Eq. \ref{Eq:Maxmag}), randomly oriented orbital inclinations lead to typically larger source-lens angular separations, and thus typically smaller lensing magnifications. Thus, at long orbital periods, lensing events are drowned-out by noise. This happens more easily for the aligned geometry, in which the secondary AGN tend to be fainter, yield lower signal-to-noise ratios, and thus require larger magnifications to be observed above the noise (see Eq.~\ref{eq:snr}).  In our simulated populations, lensing signals become undetectable above $\pobs \approx 200 \yr$ for aligned binaries, and $\approx 10^4 \yr$ for misaligned systems.  The decline in number of detectable systems at large periods is only contributed to by the decreasing probability of the lensing event occurring during the observing window.

        The overall number of long-period lensing binaries will far outnumber those with periods of a few years that allow for multiple events to be observed in $\sim$decade-long surveys.  The duration of these lensing events, proportional to the orbital period, will be months to a few years in duration.  Most of these systems will have low lensing magnifications, but as long as the damped random walk model remains applicable (which we assume in \figref{fig:long}), the intrinsic noise will be white in character, possibly making lensing flares more distinguishable.  At the same time, as this regime of long-duration AGN variability is only now being explored with precision photometry, additional confusion sources (e.g.~[partial-]tidal disruption events, changing-look/changing-character AGN, etc).


\section{Discussion}
    \label{sec:disc}

    There are no confirmed examples of sub-parsec MBH binaries, and many uncertainties regarding both their formation (e.g.~the rate and parametric distributions of MBHs in galaxy mergers) and particularly their binary evolution (e.g.~stellar scattering efficiencies, local gas densities, etc). The overall predicted occurrence rates of luminous close-binaries has systematic uncertainties that are likely at least a factor of a few.  Additionally, the binary merger physics is particularly challenging to model in extreme mass-ratio binaries, which represent a noticeable fraction of the target self-lensing population.
    
    If the signal observed in recent pulsar timing array data \citep{NANOGrav+2020_12p5gwb} is confirmed to be an astrophysical gravitational wave (GW) background, expected to be determined within the next couple of years \citep{Pol+2021}, it would definitively demonstrate the existence of a cosmological population of MBHBs.  Our populations of MBHBs derived from Illustris produce GW amplitudes roughly a factor of $\approx 2 - 3$ lower than the prospective GW signal.  If the GW signal is confirmed, this would suggest broad agreement and perhaps a higher occurrence rate of sources.  Note, however, that extreme mass-ratio binaries contribute very little to the background \citep[e.g.][]{Kelley+2018}, and also that numerous parameters in binary populations and evolution contribute to the GW background amplitude \citep[e.g.][]{Middleton+2021}.  Nearby and particularly-massive MBHBs detected through self-lensing would be promising candidates for targeted GW searches \citep[e.g.][]{Arzoumanian+2020_3C66B}, and the identification of multimessenger sources presents exciting opportunities for novel scientific studies \citep[e.g.][]{Kelley+2019b}.
    
    Based on our models, we predict that the Vera Rubin Observatory's LSST survey should detect $\sim 10 - 100$ of self-lensing MBHB systems (see Table~\ref{tab:tab}).  The uncertain distribution of alignment angles, between binary orbital planes and those of the secondary accretion disks, leads to a difference in detection rate of about a factor of seven.  For lensing signals to be detectable, the orbital plane needs to be viewed nearly edge-on by the observer to produce sufficiently small angular separations between source and lens.  If the secondary AGN disk is aligned with the orbit, it also presents a smaller projected area and thus much lower flux, making fewer systems detectable.  The relative orientation of the secondary disk also has a significant impact on the distribution of lensing magnifications, and the relative brightness of different photometric bands.  This suggests that orientations can be measured from a lensing population, the degeneracy between detection rates and disk alignment distributions can be broken, and constraints can be made on the overall occurrence rate of close-separation MBHBs. With the total binary mass also measurable from the orbital period, the binary orbital parameters (including individual masses, eccentricity, argument of pericenter, separation, and inclination) of self-lensing MBHBs can be very well constrained \citep[see][]{HuSpikey+2020}. Because these types of systems will also tend to have at least millions of years before their eventual coalescence, they will be long-term laboratories for studying the accretion and dynamics of binaries.
    
    The other most-promising indicators of AGN in binary systems are likely periodic photometric variability \citep{Graham+2015, Charisi+2016, LiuSBHB+2019, ChenDES_SBHBs+2020}, or time-variable / kinematically-offset broad emission lines \citep[e.g.][]{Eracleous+2012,Runnoe+2017}.  There are significant challenges to both methods, particularly due to contamination from noise sources in single AGN \citep[e.g.][]{Liu+2018, Kelley-2021}.  The damped random walk (DRW) red-noise that is characteristic of AGN is particularly challenging to distinguish from long-duration periodic signatures.  Accounting for DRW noise in our models decreases the number of binaries with detectable Doppler variable lightcurves by roughly a factor of four (see Tab.~\ref{tab:tab}).  Lensing flares, however, tend to have much larger amplitudes, and our models show an only mild decrease ($\approx 15\%$) in detection rate when including red noise.  We predict that both lensing flares and Doppler variations should be simultaneously detectable for a few to tens of binaries with LSST.

    The Vera Rubin Observatory's LSST is extremely promising for AGN binary observations.  According to our population models, current surveys do not have sufficient depth and coverage to plausibly expect detections.  Using simple parameterizations for Pan-STARRS \citep{LiuSBHB+2019}, PTF \citep{Charisi+2016}, Kepler \citep{KeplerSpecs:2010}, and Gaia\footnote{\href{https://www.cosmos.esa.int/web/gaia/science-performance}{https://www.cosmos.esa.int/web/gaia}}, we find zero detections within the $2-\sigma$ range; while for CRTS \citep{Graham+2015} and ZTF \citet{Bellm+2019_ZTF}, the median plus $2-\sigma$ detection rate is roughly one source after 5 years of data.  The survey parameters and resulting detection rates are shown in Tab.~\ref{tab:surveys}.
    It is worth noting that the only existing, strong candidate for a lensing MBH binary was detected by Kepler as KIC 11606854 \citep[``Spikey";][]{HuSpikey+2020, Kun+2020}.  This source shows a very short ($\sim10$ day) flare, that is well resolved and clearly above the noise.  Our models firmly predict that no lensed AGN should have been observable in the Kepler field, owing to the rarity of MBHBs and Kepler's small field of view.  Indeed, Spikey is one of only dozens of AGN in the Kepler field \citep{Smith+2018}, implying an extremely fortuitous event, a very high event rate, or an alternative source model.  If Spikey is confirmed to be a self-lensing MBHB, a detection rate of $\gtrsim 10^{-2}$ would be very difficult to explain with our Illustris-based MBHB populations.
    
    In general, bluer optical bands show higher lensing magnifications because their compact emitting regions are more uniformly magnified.  However, because AGN and particularly post galaxy-merger galactic nuclei tend to have high column densities \citep{Koss+2016, Ricci+2017, Koss+2018}, bluer near-optical observing bands may often be partially or entirely obscured.  The columns and obscuration fractions of AGN are believed to increase significantly for observers viewing AGN disks edge-on \citep[e.g.][]{Ramos-Almeida+Ricci-2017}.  If circumbinary disks tend to be aligned with binary orbits, and if these dusty torii are still present in close binaries, this could present a significant challenge to observing self-lensing signatures.  Such obscuring regions, however, may offer a different way to find reprocessed lensing and Doppler boost signatures in the IR \citep{DOrazioHaiman:2017}.  Alternatively, dusty torii that tend to be perpendicular to the binary orbital plane, as can occur for eccentric orbits \citep[e.g.][]{MartinLubow:2017}, would produce much lower obscuring fractions.
    
    Both theory \citep[e.g.][]{Netzer+Laor-1993, Nenkova+2008} and observations \citep[e.g.][]{Suganuma+2006, Koshida+2014} give typical radii of the dusty torus to be $\rdust \approx 0.1 \pc \, \lr{L_\nu / 10^{45} \, \textrm{erg s}^{-1}}^{1/2}$.  Applying this relation to the combined luminosity of both MBHs in our population of lensing-detectable binaries gives a median and interquartile range: \mbox{$\rdust = 7.7^{+5.0}_{-2.8} \E{-2} \pc = 13^{+12}_{-6} \, \sepa$}, where $\sepa$ is the binary separation.  If we assume that the dusty torus is disrupted if $\rdust / \sepa < 10$ ($\rdust / \sepa < 3$), then $38\%$ ($13\%$) of lensing binaries would be unobscured even if aligned with the circumbinary disk.  If, on the other hand, we assume that circumbinary disks are randomly oriented and obscuration is determined by the covering fraction of torii, then we might expect $\approx 25\%$ of lensing systems to be unobscured\footnote{Estimated using the obscuration/Type-2 fraction of AGN from \citet{Hasinger-2008, Merloni+2014, Suh+2019}, and X-Ray bolometric corrections from \citet{Runnoe+2017}.}.  Applying these fractions to the expected detection rates for LSST, assuming it covers $1/4$ of the sky, and assuming that obscured binaries are completely undetectable, we get a very conservative estimate of $\approx 6$ detectable systems if all are aligned, and $\approx 30$ if all are misaligned.


\section{Conclusions}
    
    In this study we have combined sophisticated populations of MBH binaries with models of AGN disks and gravitational self-lensing which will be detectable with fast-cadence all-sky surveys like VRO's LSST.  Our key conclusions are:
    \begin{itemize}
        \item We expect 10--100 self-lensing systems to be detectable with LSST in our fiducial models, even after estimating the effects of dusty-torus obscuration and including intrinsic AGN variability.
        \item Most current surveys are unlikely to have sufficient coverage to contain detectable lensing events\footnote{Our models do show CRTS and ZTF making a single detection in $\approx2\%$ of realizations, with 5 years of data}.
        \item Detectable lensing events have median magnifications of $\sim100\%$, with durations of a few tens of days.  Sources typically have total masses between $10^8$--$10^9\msol$, mass-ratios $10^{-2}$--$10^{-1}$, and orbital periods of multiple years.
        \item Because lensing events are intrinsically rare, surveys with a given depth should attempt to cover as large an area of the sky as possible with a cadence of 3-7 days.  Lensing events can be chromatic for some configurations, brightening more in bluer bands, but those bands may also be subject to additional obscuration.
        \item Detectable lensing signatures require viewing angles near the orbital plane.  A key uncertainty in our models is whether the secondary accretion disk tends to be aligned with the orbit, and thus typically also edge-on, or may be misaligned and randomly oriented.
        \item Magnifications tend to be nearly achromatic, but differences between observing bands can strongly probe the AGN disk inclination and structure.  Peak magnifications tend to closely match the point-source approximation, with detectable deviations only in misaligned geometries.
        \item We find that intrinsic DRW red-noise in AGN does not significantly inhibit lensing detections, but does make coincident Doppler variations much more difficult. Doppler signals should be detectable in $\approx 1/3$ of lensing binaries.
        \item Our fiducial detection models require short enough orbital periods to observe multiple, periodic lensing flares.  There should, however, be orders of magnitude more lensing binaries at much longer orbital periods ($10$s--$1000$s yr) having flares that last months to years.
    \end{itemize}

    The modeling of MBH binary populations includes numerous significant uncertainties, both regarding their rates of formation and the complex physics of their binary evolution.  Our understanding of MBH binaries interacting with circumbinary accretion flows has rapidly progressed in the last decade, but remains very challenging to probe with detailed simulations.  The efficiency of driving material to accrete onto each component MBH is particularly important for predicting and understanding the resulting electromagnetic signatures.  Further modeling of circumbinary accretion is thus particularly important, especially including effects such as AGN feedback and complex geometries.
    
    The confident detection of self-lensing in an AGN presents many exciting opportunities. Foremost, it would be the first confirmed detection of a gravitationally-bound MBH binary.  Measurement of the orbital period and lensing characteristics can strongly constrain the MBH components, their accretion, and their orbital parameters. As no MBHBs have yet been observationally confirmed, binary populations remain uncertain.  The detection of an ensemble of self-lensing systems could place tight constraints on MBH binary formation and evolution.  In exceptional systems with high signal to noise ratios and small impact parameters, lensing events could be used to probe the inner structure of AGN accretion disks.  Nearby and high mass-ratio systems could also be promising multi-messenger sources combined with low-frequency gravitational waves which are expected to be detected within the next few years.


\section*{Acknowledgements}
    \vspace{-6pt}

    We greatly appreciate helpful discussion and feedback from Maria Charisi, \cafg, and the NANOGrav Astrophysics Working Group.

    This research made use of \astropy, a community-developed core Python package for Astronomy \citep{astropy2013}, in addition to \scipy~\citep{scipy}, \ipython~\citep{ipython}, \jupyter~notebook~\citep{jupyter}, \& \numpy~\citep{numpy2011}.  All figures were generated using \matplotlib~\citep{matplotlib2007}.  Kernel density estimation was performed using the \kalepy{} package (\href{https://github.com/lzkelley/kalepy}{github.com/lzkelley/kalepy}) \citep{kalepy2021}.
    
    DJD received funding from the European Union's Horizon 2020 research and innovation programme under the Marie Sklodowska-Curie grant agreement No. 101029157 and through Villum Fonden grant No. 29466.  
    
    RD was supported in part by NASA grant GO0-21087A.

\section*{Data Availability}
\vspace{-6pt}

    The Illustris data is available online at \href{https://www.illustris-project.org/}{www.illustris-project.org} \citep{nelson2015}, and Illustris-TNG data at \href{https://www.tng-project.org/}{www.tng-project.org} \citep{Nelson+1812.05609}.  One realization of our full binary population is available at \href{https://zenodo.org/record/4068485}{zenodo:4068485} \citep{Kelley-2020_kde-data}.



\bibliographystyle{mnras}
\bibliography{refs}


\appendix

\section{Supplementary Material}

    \begin{figure}
        \centering
        \includegraphics[width=\columnwidth]{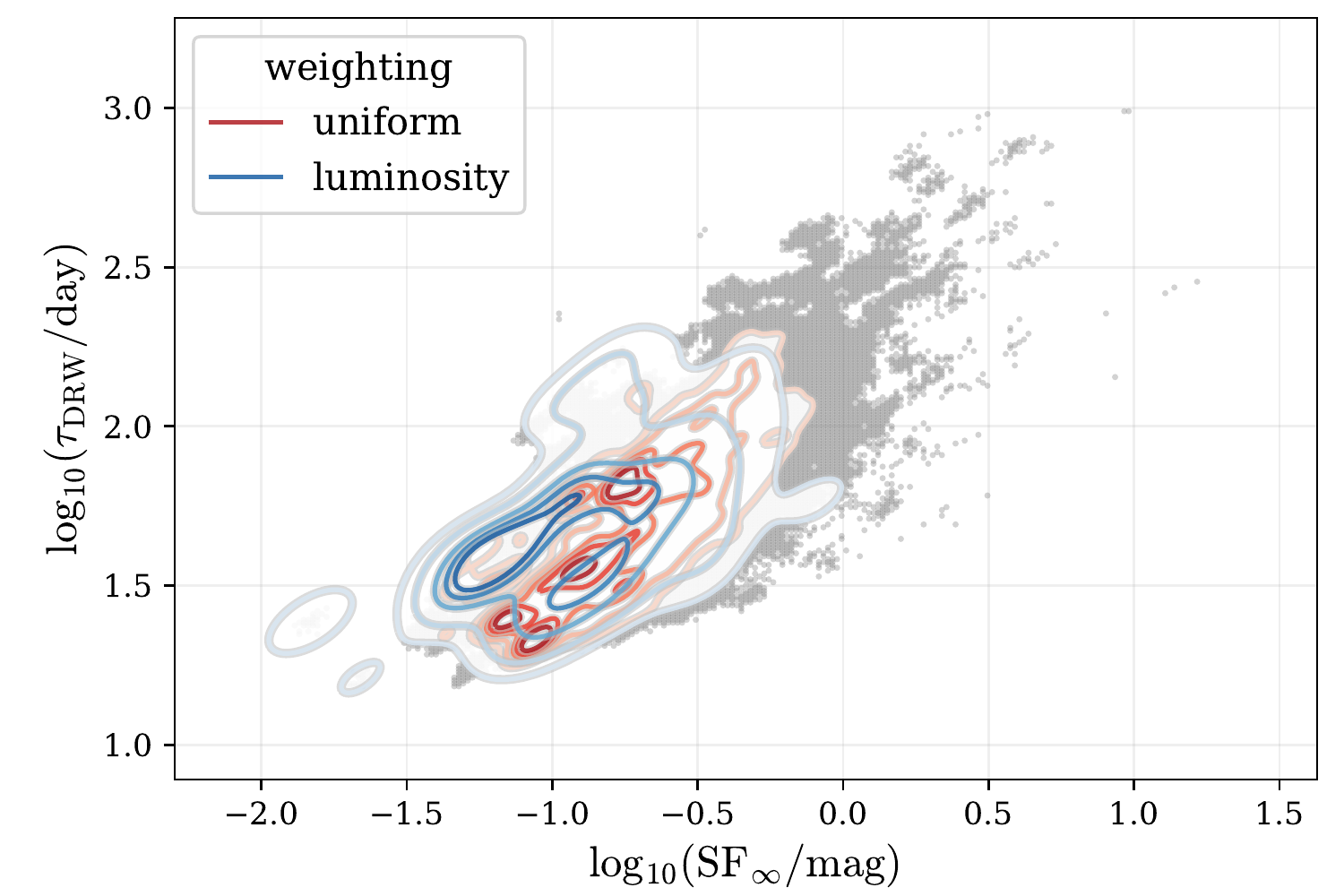}
        \caption{\textbf{Damped Random Walk (DRW) parameters for the secondary MBH}: contours are shown for both \textcolor{mplred}{uniform weightings (red)} and systems \textcolor{mplblue}{weighted by their luminosity (blue)}.  The contours correspond to the quantiles: $\{0.10, 0.25, 0.50, 0.75, 0.90\}$.}
        \label{fig:app_drw}
    \end{figure}

    \begin{figure*}
        \centering
        \includegraphics[width=1.0\textwidth, height=280px]{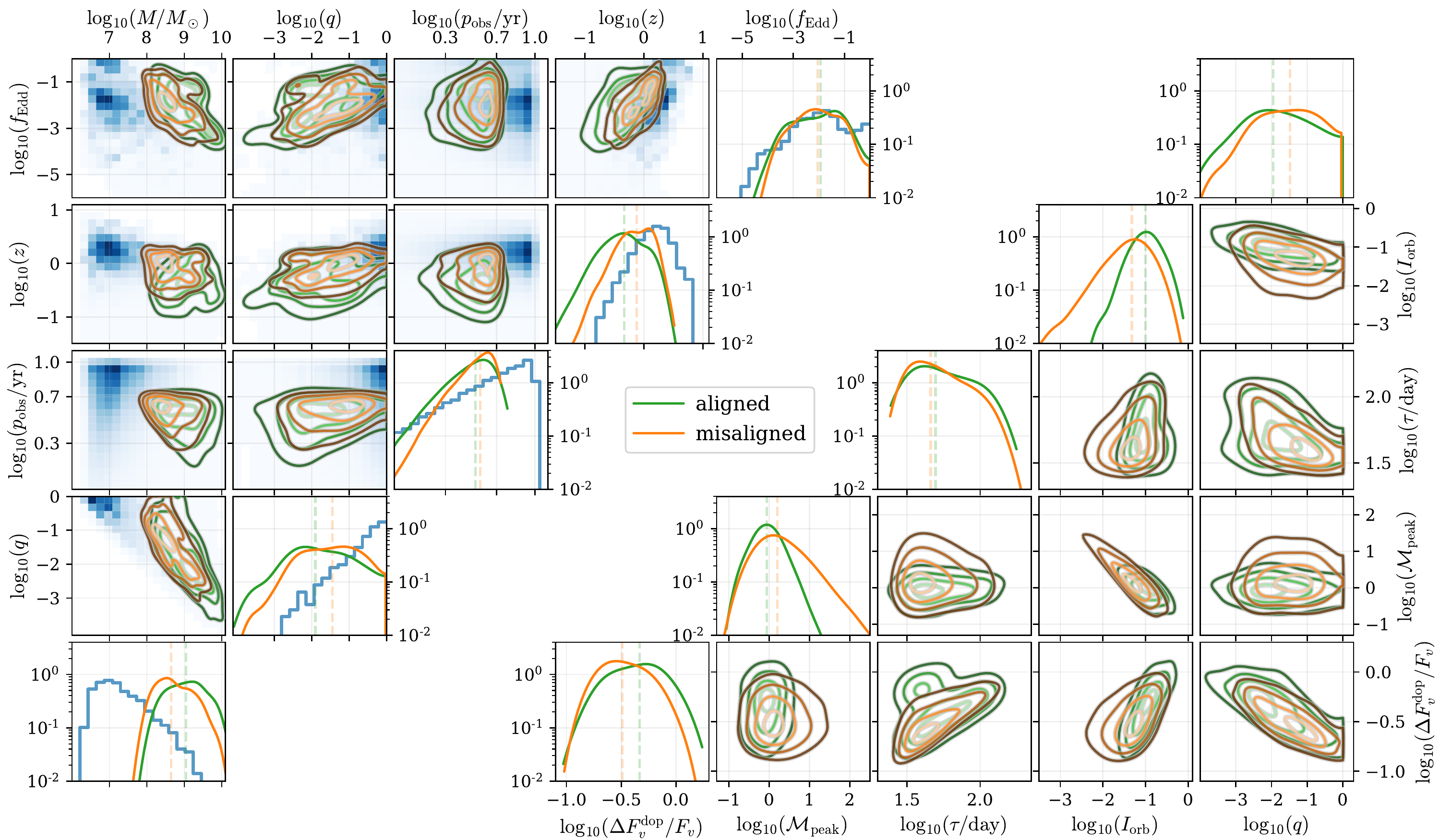}
        \vspace{-16pt}
        \caption{\textbf{Binary population parameters}: \textcolor{mplblue}{all simulated binaries (blue; $\approx 1.5\E{6}$)} compared to those that are \textcolor{mplorange}{lensing-detectable aligned systems (orange)} and \textcolor{mplgreen}{lensing-detectable misaligned systems (green)}.  `Aligned' systems are those having circum-single disks coaligned with the binary orbital plane.  All distributions are probability densities.  The detection criteria are: a flux above $5\E{-30} \, \fluxunits{}$ in the R-Band, a lensing duration longer than $30$ days, an observed orbital period shorter than $5$ years, and a peak lensing magnitude larger than $\mathrm{SNR}^{-1} + 0.05$.  Lensing systems have large total masses, but relatively extreme mass ratios peaking near $q \sim 10^{-2}$.  The accretion partition function, which gives the majority of mass-accretion to the secondary, allows for small overall Eddington ratios, while the Eddington ratio of the secondaries (and thus their luminosities) remain high.
        }
        \label{fig:pop}
    \end{figure*}

    \begin{table*}
        \centering
        \setlength{\tabcolsep}{3pt}
        \renewcommand{\arraystretch}{1.6}
        \begin{tabular}{c c | c c | c c | c c}
               \multirow{2}{*}{\shortstack{sensitivity\\$[\fluxunits{}]$}} &    & \multicolumn{2}{c}{lensing}   & \multicolumn{2}{c}{Doppler} & \multicolumn{2}{c}{both} \\
             & period & white & DRW & white & DRW & white & DRW \\
            \hline
         
            \multirow{2}{*}{$3\E{-30}$} & $3.0$ yr & $25^{+8}_{-11} \, \lr{130^{+22}_{-20}}$  & $22^{+10}_{-9.0} \, \lr{120^{+22}_{-21}}$  & $110^{+20}_{-20} \, \lr{320^{+30}_{-30}}$  & $87^{+17}_{-17} \, \lr{210^{+30}_{-21}}$  & $17^{+8.0}_{-8.0} \, \lr{80^{+18}_{-20}}$  & $13^{+10}_{-6.0} \, \lr{59^{+17}_{-13}}$   \\
            & $5.0$ yr & $73^{+16}_{-13} \, \lr{510^{+31}_{-44}}$  & $60^{+15}_{-13} \, \lr{450^{+32}_{-38}}$  & $220^{+31}_{-18} \, \lr{670^{+51}_{-38}}$  & $160^{+23}_{-20} \, \lr{370^{+36}_{-32}}$  & $44^{+15}_{-10} \, \lr{250^{+32}_{-26}}$  & $26^{+12}_{-9.0} \, \lr{130^{+23}_{-28}}$   \\
            \rule{0pt}{16pt}

            \multirow{2}{*}{$3\E{-29}$} & $3.0$ yr & $1.0^{+2.0}_{-1.0} \, \lr{13^{+8}_{-6}}$  & $1.0^{+2.0}_{-1.0} \, \lr{12^{+7}_{-6}}$  & $5.0^{+6.0}_{-4.0} \, \lr{20^{+11}_{-7}}$  & $4.0^{+5.0}_{-3.0} \, \lr{14^{+9}_{-6}}$  & $0.0^{+2.0}_{-0.0} \, \lr{5.0^{+5.7}_{-3.0}}$  & $0.0^{+1.0}_{-0.0} \, \lr{4.0^{+4.0}_{-3.0}}$   \\
            & $5.0$ yr & $3.0^{+3.0}_{-2.7} \, \lr{50^{+11}_{-12}}$  & $2.0^{+4.0}_{-2.0} \, \lr{47^{+10}_{-12}}$  & $10^{+7.0}_{-6.0} \, \lr{44^{+13}_{-12}}$  & $6.0^{+6.0}_{-4.0} \, \lr{24^{+12}_{-9.0}}$  & $1.0^{+3.0}_{-1.0} \, \lr{16^{+10}_{-6.0}}$  & $1.0^{+1.0}_{-1.0} \, \lr{9.0^{+6.0}_{-5.7}}$

        \end{tabular}
        \caption{\textbf{All sky detection rates for different signals and survey parameter combinations.}  The signatures are lensing peaks, Doppler modulations, and the combination of the two (both).  For each signal we compare between a white-noise only model, and also including Damped Random Walk (DRW) variations which depend on the duration of a prospective signal.  The first number shown in each cell assumes the secondary AGN disk is aligned with the orbital plane, while the number in parenthesis assumes a random orientation.  These values do not take into account obscuration (see Sec.~\ref{sec:disc}).  Uncertainties are $2-\sigma$ reflecting Poisson-like variations across 100 realizations of our binary populations with fixed parameters.}
        \label{tab:tab}
    \end{table*}

    \begin{table*}
        \renewcommand{\arraystretch}{1.4}
        \centering
        \begin{tabular}{c | c | c | c | c | c }
            Survey      & $F_{\nu,\mathrm{sens}} [\fluxunits]$  & Coverage $[str./(4\pi)]$  & Duration [yrs]   & Cadence [days]     & Detections [median$^{+2\sigma}_{-2\sigma}$] \\ \hline
            LSST        & $3\E{-30}$                            & 0.8                       & 10               & 3                  & $60_{47}^{75}$    \\
            CRTS        & $4\E{-28}$                            & 0.8                       & 10               & 7                  & $0_{0}^{0.8}$     \\
            PTF         & $2\E{-28}$                            & 0.066                     & 3.75             & 5                  & $0_{0}^{0}$       \\
            ZTF         & $2\E{-28}$                            & 0.8                       & 5                & 3                  & $0_{0}^{0.75}$    \\
            Pan-STARRS  & $4\E{-29}$                            & $2\E{-4}$                 & 5                & 3                  & $0_{0}^{0}$       \\
            Kepler      & $10^{-27}$                            & $3\E{-3}$                 & 3.5              & 0.25               & $0_{0}^{0}$       \\
            Gaia      & $2\E{-28}$                              & 1.0                       & 10               & 30                 & $0_{0}^{0}$       \\
            \hline
        \end{tabular}
        \caption{Adopted survey parameters and number of expected detections.  The detection rates are calculate over 100 realizations of our binary populations.}
        \label{tab:surveys}
    \end{table*}
    
\label{lastpage}

\end{document}